\documentclass{PoS}

\def\Slash#1{\not\!\!#1}

\title{Analytical relation between the Polyakov loop and 
Dirac eigenvalues in temporally odd-number lattice QCD}

\ShortTitle{Analytical relation between Polyakov loop and
Dirac eigenvalues in odd-$N_t$ lattice QCD}

\author{\speaker{Hideo Suganuma}, Takahiro M. Doi \\
        Department of Physics \& Division of Physics and Astronomy, 
Graduate School of Science, \\
Kyoto University, 
Kitashirakawaoiwake, Sakyo, Kyoto 606-8502, Japan\\
        E-mail: \email{suganuma@scphys.kyoto-u.ac.jp}}

\author{Takumi Iritani \\
High Energy Accelerator Research Organization (KEK), 
Tsukuba, Ibaraki 305-0801, Japan}

\abstract{
We derive an analytical gauge-invariant relation between the Polyakov loop 
$\langle L_P \rangle$ and the Dirac eigenvalues $\lambda_n$ in QCD, i.e., 
$\langle L_P \rangle \propto 
\sum_n \lambda_n^{N_t -1} \langle n|\hat U_4|n \rangle$, 
on a temporally odd-number lattice, where the temporal lattice size $N_t$ 
is odd. Here, we use an ordinary square lattice with 
the normal 
\break
(nontwisted) periodic boundary condition for link-variables 
in the temporal direction. 
This relation is a Dirac spectral representation of 
the Polyakov loop in terms of Dirac eigenmodes $|n\rangle$.
Because of the factor $\lambda_n^{N_t -1}$ in the Dirac spectral sum, 
this analytical relation indicates 
negligibly small contribution of 
low-lying Dirac modes to the Polyakov loop 
in both confined and deconfined phases, 
while the low-lying Dirac modes are essential for chiral symmetry breaking. 
Also, we numerically confirm the analytical relation, 
non-zero finiteness of $\langle n|\hat U_4|n \rangle$, 
and tiny contribution of low-lying Dirac modes to the Polyakov loop 
in lattice QCD simulations. 
Thus, we conclude that low-lying Dirac modes 
are not essential modes for confinement, 
and there is no direct one-to-one correspondence between 
confinement and chiral symmetry breaking in QCD.
}

\FullConference{
The 31st International Symposium on Lattice Field Theory - LATTICE 2013\\
		July 29 - August 3, 2013\\
		Mainz, Germany}

\begin{document}

\section{Introduction: color confinement and chiral symmetry breaking}

Quantum chromodynamics (QCD) has two outstanding 
nonperturbative phenomena of spontaneous chiral-symmetry breaking \cite{NJL61}
and color confinement \cite{N74tH81,YS0809}.
However, their relation is not yet known directly from QCD, 
and to clarify their precise relation is one of the important problems 
in theoretical physics \cite{SST95,M95W95,G06BGH07,GIS12,S111213,IS13}. 

For chiral symmetry breaking, 
the standard order parameter 
is the chiral condensate $\langle \bar qq \rangle$,
and low-lying Dirac modes play the essential role, 
as is indicated by the Banks-Casher relation \cite{BC80}.
For quark confinement, 
the Polyakov loop $\langle L_P \rangle$ is 
one of the typical order parameters, and 
it relates to the single-quark free energy $E_q$ as 
$\langle L_P \rangle \propto e^{-E_q/T}$ 
at temperature $T$. 
The Polyakov loop is the order parameter 
of spontaneous breaking of the $Z_{N_c}$ center symmetry 
in QCD \cite{Rothe12}.

A strong correlation between confinement and chiral symmetry breaking 
has been suggested by the simultaneous phase transitions of deconfinement and 
chiral restoration in lattice QCD both 
at finite temperatures and in a finite-volume box \cite{Rothe12}.
Their strong correlation has been also suggested 
in terms of QCD-monopoles \cite{SST95,M95W95}, 
which topologically appear in QCD in the maximally Abelian gauge 
\cite{N74tH81}. 
Actually, by removing the monopoles from the QCD vacuum, 
confinement and chiral symmetry breaking are 
simultaneously lost in lattice QCD \cite{M95W95},
which indicates an important role of QCD-monopoles 
to both confinement and chiral symmetry breaking, and thus 
these two phenomena seem to be related via the monopole.

In our previous studies \cite{GIS12,S111213,IS13}, 
aiming to know the direct relation between 
confinement and chiral symmetry breaking, 
we study confinement in terms of Dirac eigenmodes in QCD,
because low-lying Dirac modes are essential 
for chiral symmetry breaking \cite{BC80}.
Based on completeness of the Dirac-mode basis, 
we consider Dirac-mode expansion and projection 
where the Dirac-mode space is restricted, 
and investigate the role of low-lying Dirac modes 
to confinement in SU(3) lattice QCD 
\cite{GIS12,S111213,IS13}.
Remarkably, 
even after removing the coupling to the low-lying Dirac modes, 
the following numerically lattice-QCD results are obtained: 
\begin{enumerate}
\item The Wilson loop obeys the area law \cite{GIS12,S111213}, 
which indicates quark confinement. 
\item The string tension, i.e., the confining force, 
is almost unchanged \cite{GIS12,S111213}.
\item The Polyakov loop remains to be almost zero \cite{IS13}, 
indicating $Z_3$-unbroken confinement phase.
\end{enumerate}
In fact, quark confinement is kept 
even in the absence of the low-lying Dirac modes. 

In this study, we derive an analytical relation between 
the Polyakov loop and the Dirac modes in temporally odd-number lattice QCD, 
where the temporal lattice size is odd, 
and discuss the relation between confinement and chiral symmetry breaking.

\section{Lattice QCD formalism: general arguments}

To begin with, we state the setup condition of 
lattice QCD formalism adopted in this study.
We use an ordinary square lattice with spacing $a$ and 
size $N_s^3 \times N_t$.
The normal (nontwisted) periodic boundary condition is used 
for the link-variable $U_\mu(s)={\rm e}^{iagA_\mu(s)}$ 
in the temporal direction, 
with the gluon field $A_\mu(s)$, the gauge coupling $g$ and the site $s$.
%
(This temporal periodicity is physically required at finite temperature.)
In this paper, we take SU($N_c$) ($N_c$: color number) 
as the gauge group of the theory. 
However, arbitrary gauge group $G$ can be taken 
for most arguments in the following. 

\subsection{Dirac operator, Dirac eigenvalues and Dirac modes in lattice QCD}

In lattice QCD, the Dirac operator 
$\Slash D = \gamma_\mu D_\mu$ is expressed with 
$U_\mu(s)={\rm e}^{iagA_\mu(s)}$ as
\begin{eqnarray}
 \Slash{D}_{s,s'} 
 \equiv \frac{1}{2a} \sum_{\mu=1}^4 \gamma_\mu 
\left[ U_\mu(s) \delta_{s+\hat{\mu},s'}
 - U_{-\mu}(s) \delta_{s-\hat{\mu},s'} \right],
\end{eqnarray}
where $\hat\mu$ is the unit vector in $\mu$-direction in the lattice unit, 
and $U_{-\mu}(s)\equiv U^\dagger_\mu(s-\hat \mu)$.
Adopting hermitian $\gamma$-matrices as $\gamma_\mu^\dagger=\gamma_\mu$, 
the Dirac operator $\Slash D$ is anti-hermitian and satisfies 
$\Slash D_{s',s}^\dagger=-\Slash D_{s,s'}$.
We introduce the normalized Dirac eigen-state $|n \rangle$ as 
\begin{eqnarray}
\Slash D |n\rangle =i\lambda_n |n \rangle, \qquad
\langle m|n\rangle=\delta_{mn}, 
\end{eqnarray}
with the Dirac eigenvalue $i\lambda_n$ ($\lambda_n \in {\bf R}$).
Because of $\{\gamma_5,\Slash D\}=0$, the state 
$\gamma_5 |n\rangle$ is also an eigen-state of $\Slash D$ with the 
eigenvalue $-i\lambda_n$. 
Here, the Dirac eigen-state $|n \rangle$ 
satisfies the completeness of 
\begin{eqnarray}
\sum_n |n \rangle \langle n|=1.
\end{eqnarray}
For the Dirac eigenfunction $\psi_n(s)\equiv\langle s|n \rangle$, 
the explicit form of the Dirac eigenvalue equation \\
$\Slash D \psi_n(s)=i\lambda_n \psi_n(s)$ 
in lattice QCD is written by 
\begin{eqnarray}
\frac{1}{2a} \sum_{\mu=1}^4 \gamma_\mu
[U_\mu(s)\psi_n(s+\hat \mu)-U_{-\mu}(s)\psi_n(s-\hat \mu)]
=i\lambda_n \psi_n(s).
\end{eqnarray}
The Dirac eigenfunction $\psi_n(s)$ can be 
numerically obtained in lattice QCD, besides a phase factor. 
By the gauge transformation of 
$U_\mu(s) \rightarrow V(s) U_\mu(s) V^\dagger (s+\hat\mu)$, 
$\psi_n(s)$ is gauge-transformed as 
\begin{eqnarray}
\psi_n(s)\rightarrow V(s) \psi_n(s),
\label{eq:GTDwf}
\end{eqnarray}
which is the same as that of the quark field, although, 
to be strict, there can appear an irrelevant $n$-dependent 
global phase factor $e^{i\varphi_n[V]}$, 
according to arbitrariness of the phase in 
the basis $|n \rangle$
\cite{GIS12}.

Note that the spectral density $\rho(\lambda)$ 
of the Dirac operator $\Slash D$ relates to chiral symmetry breaking.
For example, from Banks-Casher's relation \cite{BC80}, 
the zero-eigenvalue density $\rho(0)$ leads to 
$\langle\bar qq \rangle$ as 
\begin{eqnarray}
\langle \bar qq \rangle=-\lim_{m \to 0} \lim_{V_{phys} \to \infty} 
\pi\rho(0), \qquad
\rho(\lambda)\equiv 
\frac{1}{V_{phys}}\sum_{n}\langle \delta(\lambda-\lambda_n)\rangle,
\end{eqnarray}
with space-time volume $V_{phys}$.
Thus, the low-lying Dirac modes can be regarded as the essential modes 
responsible to spontaneous chiral-symmetry breaking in QCD.

\subsection{Operator formalism in lattice QCD}

Now, we present the operator formalism 
in lattice QCD \cite{GIS12,S111213,IS13}. 
To begin with, we introduce the link-variable operator $\hat U_{\pm \mu}$ 
defined by the matrix element of 
\begin{eqnarray}
\langle s |\hat U_{\pm \mu}|s' \rangle 
=U_{\pm \mu}(s)\delta_{s\pm \hat \mu,s'}.
\end{eqnarray}
With the link-variable operator, 
the Dirac operator and covariant derivative 
are simply expressed as 
\begin{eqnarray}
\Slash{\hat D}
=\frac{1}{2a}\sum_{\mu=1}^{4} \gamma_\mu (\hat U_\mu-\hat U_{-\mu}),
\qquad 
\hat D_\mu=\frac{1}{2a}(\hat U_\mu-\hat U_{-\mu}).
\label{eq:Dop}
\end{eqnarray}
The Polyakov loop is also simply written as the functional trace of 
$\hat U_4^{N_t}$,
\begin{eqnarray}
\langle L_P \rangle
=\frac{1}{N_c V} \langle {\rm Tr}_c \{\hat U_4^{N_t}\}\rangle
=\frac{1}{N_c V}\langle 
\sum_s {\rm tr}_c \left(\prod_{n=0}^{N_t-1} U_4(s+n\hat t)\right)\rangle,
\label{eq:PL}
\end{eqnarray}
with the four-dimensional lattice volume $V \equiv N_s^3 \times N_t$ and 
$\hat t=\hat 4$.
Here, ``${\rm Tr}_c$'' denotes the functional trace 
of ${\rm Tr}_c \equiv \sum_s {\rm tr}_c$ with 
the trace ${\rm tr}_c$ over color index.

The Dirac-mode matrix element of the link-variable operator 
$\hat U_{\mu}$ can be expressed with $\psi_n(s)$:
\begin{eqnarray}
\langle m|\hat U_{\mu}|n \rangle=\sum_s\langle m|s \rangle 
\langle s|\hat U_{\mu}|s+\hat \mu \rangle \langle s+\hat \mu|n\rangle
=\sum_s \psi_m^\dagger(s) U_\mu(s)\psi_n(s+\hat \mu).
\end{eqnarray}
Note that the matrix element is gauge invariant, 
apart from an irrelevant phase factor.
Actually, using the gauge transformation (\ref{eq:GTDwf}), we find 
the gauge transformation of the matrix element as \cite{GIS12}
\begin{eqnarray}
\langle m|\hat U_\mu|n \rangle
&=&\sum_s \psi^\dagger_m(s)U_\mu(s)\psi_n(s+\hat\mu) 
\rightarrow
\sum_s\psi^\dagger_m(s)V^\dagger(s)\cdot V(s)U_\mu(s)V^\dagger(s+\hat \mu)
\cdot V(s+\hat \mu)\psi_n(s+\hat \mu) \nonumber\\
&=&\sum_s\psi_m^\dagger(s)U_\mu(s)\psi_n(s+\hat \mu)
=\langle m|\hat U_\mu|n\rangle.
\end{eqnarray}
To be strict, there appears an $n$-dependent global phase factor, 
corresponding to the arbitrariness of the phase in the basis 
$|n \rangle$. However, this phase factor cancels 
as $e^{i\varphi_n} e^{-i\varphi_n}=1$ 
between $|n \rangle$ and $\langle n |$, and does not appear 
for physical quantities such as the Wilson loop and the Polyakov loop 
\cite{GIS12}.

\section{Analytical relation between the Polyakov loop and Dirac modes 
in temporally odd-number lattice QCD}

Now, we consider a temporally odd-number lattice, 
where the temporal lattice size $N_t$ is odd. 
Here, we use an ordinary square lattice 
with the normal (nontwisted) periodic boundary condition 
for the link-variable in the temporal direction. 
The spatial lattice size $N_s$ is taken to be larger than $N_t$, 
i.e., $N_s > N_t$. 
Note that, in the continuum limit of $a \rightarrow 0$ and 
$N_t \rightarrow \infty$, 
any number of large $N_t$ gives the same physical result.
Then, in principle, it is no problem to use the odd-number lattice.

\begin{figure}[h]
\begin{center}
\includegraphics[scale=0.3]{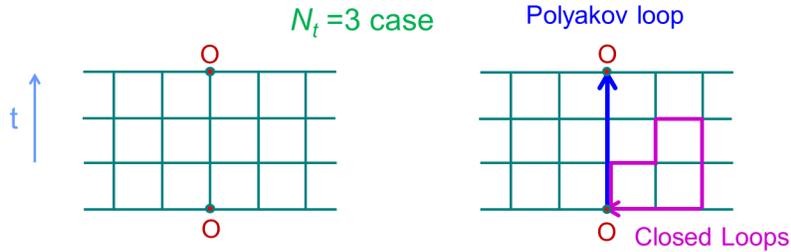}
\caption{
An example 
of the temporally odd-number lattice.
Only gauge-invariant quantities such as 
closed loops and the Polyakov loop survive in QCD.
Closed loops have even-number links on the square lattice.
}
\end{center}
\vspace{-0.3cm}
\end{figure}

In general, only gauge-invariant quantities 
such as closed loops and the Polyakov loop 
survive in QCD, according to the Elitzur theorem \cite{Rothe12}.
All the non-closed lines are gauge-variant 
and their expectation values are zero.
Note here that any closed loop (except for the Polyakov loop) 
needs even-number link-variables on the square lattice.
 (See Fig.1.)

On the temporally odd-number lattice, 
we consider the functional trace of 
\begin{eqnarray}
I\equiv {\rm Tr}_{c,\gamma} (\hat{U}_4\hat{\Slash{D}}^{N_t-1}), 
\label{eq:FT}
\end{eqnarray}
where 
${\rm Tr}_{c,\gamma}\equiv \sum_s {\rm tr}_c 
{\rm tr}_\gamma$ includes 
${\rm tr}_c$ 
and the trace ${\rm tr}_\gamma$ over spinor index.
Its expectation value 
\begin{eqnarray}
 \langle I\rangle=\langle {\rm Tr}_{c,\gamma} (\hat{U}_4\hat{\Slash{D}}^{N_t-1})\rangle 
\label{eq:FTV}
\end{eqnarray}
is obtained as the gauge-configuration average in lattice QCD.
In the case of enough large volume $V$, one can expect 
$\langle O \rangle \simeq {\rm Tr}~O/{\rm Tr}~1$ 
for any operator $O$ at each gauge configuration.

From Eq.(\ref{eq:Dop}), 
$\hat U_4\Slash{\hat D}^{N_t-1}$ 
is expressed as a sum of products of $N_t$ link-variable operators, 
since the Dirac operator $\Slash{\hat D}$ 
includes one link-variable operator in each direction of $\pm \mu$.
Then, $\hat U_4\Slash{\hat D}^{N_t-1}$ 
includes many trajectories with the total length $N_t$ in the lattice unit 
on the square lattice, as shown in Fig.2.
Note that all the trajectories with the odd-number length $N_t$ 
cannot form a closed loop 
on the square lattice, and thus give gauge-variant contribution, 
except for the Polyakov loop.

\begin{figure}[h]
\begin{center}
\includegraphics[scale=0.3]{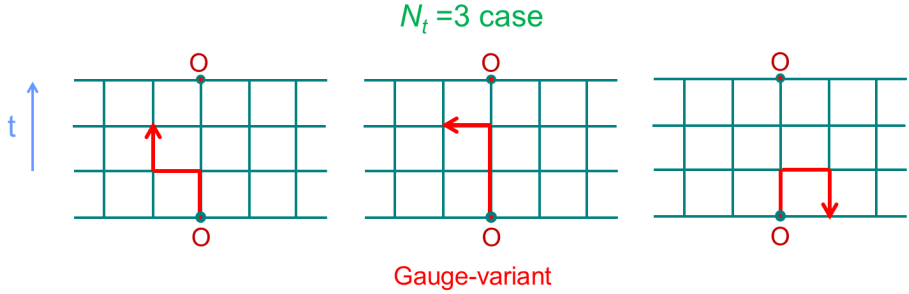}
\vspace{-0.1cm}
\caption{
A part of the trajectories stemming from 
$\langle {\rm Tr}_{c,\gamma}(\hat U_4\Slash{\hat D}^{N_t-1})\rangle$. 
For each trajectory, the first step is positive 
temporal direction, 
and the total length is 
$N_t$.
All the trajectories with the odd-number length $N_t$ 
cannot form a closed loop on the square lattice,
and thus gauge-variant, except for the Polyakov loop.
}
\end{center}
\vspace{-0.2cm}
\end{figure}

Hence, among the trajectories stemming from 
$\langle {\rm Tr}_{c,\gamma}(\hat U_4\Slash{\hat D}^{N_t-1}) \rangle$, 
all the non-loop trajectories are gauge-variant and give no contribution, 
according to the Elitzur theorem.
Only the exception is the Polyakov loop. (See Figs.2 and 3.)
[For each trajectory in $\hat U_4\Slash{\hat D}^{N_t-1}$, 
the first step is positive 
temporal direction corresponding to $\hat U_4$,
so that $\langle {\rm Tr}_{c,\gamma}(\hat U_4\Slash{\hat D}^{N_t-1})\rangle$ 
cannot include the anti-Polyakov loop $\langle L_P^\dagger \rangle$.]
Thus, in the functional trace 
$\langle I \rangle
=\langle{\rm Tr}_{c,\gamma}(\hat U_4\Slash{\hat D}^{N_t-1})\rangle$, 
only the Polyakov-loop ingredient can survive 
as the gauge-invariant quantity, and 
$\langle I \rangle$ is proportional to the Polyakov loop $\langle L_P \rangle$.

\begin{figure}[h]
\vspace{-0.3cm}
\begin{center}
\includegraphics[scale=0.3]{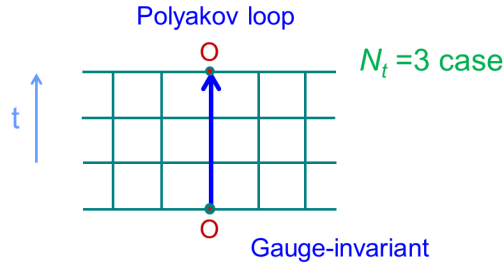}
\vspace{-0.1cm}
\caption{
Among the trajectories stemming from 
$\langle {\rm Tr}_{c,\gamma}(\hat U_4\Slash{\hat D}^{N_t-1}) \rangle$, 
only the Polyakov-loop ingredient can survive 
as the gauge-invariant quantity. 
Owing to the first factor $\hat U_4$, 
$\langle {\rm Tr}_{c,\gamma}(\hat U_4\Slash{\hat D}^{N_t-1}) \rangle$ 
does not include $\langle L_P^\dagger \rangle$.
}
\end{center}
\vspace{-0.5cm}
\end{figure}

Actually, we can mathematically derive the following relation:
\begin{eqnarray}
\langle I\rangle
&=&\langle {\rm Tr}_{c,\gamma} (\hat U_4 \hat{\Slash D}^{N_t-1}) \rangle
\nonumber \\
&=&\langle {\rm Tr}_{c,\gamma} \{\hat U_4 (\gamma_4 \hat D_4)^{N_t-1}\} \rangle
\quad \qquad \qquad ~~~~~{\rm 
(
\raisebox{1.2ex}{.}\raisebox{.2ex}{.}\raisebox{1.2ex}{.} 
~only~gauge\hbox{-}invariant~terms~survive)} 
\nonumber \\
&=&4\langle {\rm Tr}_{c} (\hat U_4 \hat D_4^{N_t-1}) \rangle
\quad \qquad \qquad \qquad ~~~~~~(
\raisebox{1.2ex}{.}\raisebox{.2ex}{.}\raisebox{1.2ex}{.} 
~\gamma_4^{N_t-1}={1}, 
~{\rm tr}_\gamma {1}=4) 
\nonumber \\
&=&\frac{4}{(2a)^{N_t-1}}
\langle {\rm Tr}_{c} \{\hat U_4 (\hat U_4-\hat U_{-4})^{N_t-1}\} \rangle
\quad ~~(
\raisebox{1.2ex}{.}\raisebox{.2ex}{.}\raisebox{1.2ex}{.} 
~\hat D_4=\frac{1}{2a}(\hat U_4-\hat U_{-4}))
\nonumber \\
&=&\frac{4}{(2a)^{N_t-1}} \langle {\rm Tr}_{c} \{ \hat U_4^{N_t} \}\rangle
\qquad \qquad \qquad ~~~~~{\rm 
(\raisebox{1.2ex}{.}\raisebox{.2ex}{.}\raisebox{1.2ex}{.} 
~only~gauge\hbox{-}invariant~terms~survive)} 
\nonumber \\
&=&\frac{12V}{(2a)^{N_t-1}}\langle L_P \rangle. 
\label{eq:FTdetail}
\end{eqnarray}
We thus obtain the relation between 
$\langle I\rangle = \langle {\rm Tr}_{c,\gamma}
 (\hat U_4 \hat{\Slash D}^{N_t-1}) \rangle$ 
and the Polyakov loop $\langle L_P \rangle$,
\begin{eqnarray}
\langle I\rangle
=\langle {\rm Tr}_{c,\gamma} (\hat U_4 \hat{\Slash D}^{N_t-1}) \rangle
=\frac{12V}{(2a)^{N_t-1}}\langle L_P \rangle. 
\label{eq:FTtoPL}
\end{eqnarray}

On the other hand, we calculate the functional trace 
in Eq.(\ref{eq:FTV}) using the complete set of 
the Dirac-mode basis $|n\rangle$ satisfying $\sum_n |n\rangle \langle n|=1$, 
and find the Dirac-mode representation of 
\begin{eqnarray}
 \langle I\rangle=\sum_n\langle n|\hat{U}_4\Slash{\hat{D}}^{N_t-1}|n\rangle
=i^{N_t-1}\sum_n\lambda_n^{N_t-1}\langle n|\hat{U}_4| n \rangle. 
\label{eq:FTtoD}
\end{eqnarray}
Combing Eqs.(\ref{eq:FTtoPL}) and (\ref{eq:FTtoD}), we obtain the analytical 
relation between the Polyakov loop $ \langle L_P \rangle$ 
and the Dirac eigenvalues $i\lambda_n$: 
\begin{eqnarray}
\langle L_P \rangle=\frac{(2ai)^{N_t-1}}{12V}
\sum_n\lambda_n^{N_t-1}\langle n|\hat{U}_4| n \rangle. 
\label{eq:PLvsD}
\end{eqnarray}
This is a direct relation between the Polyakov loop $\langle L_P\rangle$ 
and the Dirac modes in QCD, and is 
\break
mathematically valid on the temporally odd-number lattice 
in both confined and deconfined phases. 
The relation (\ref{eq:PLvsD}) is 
a Dirac spectral representation of the Polyakov loop, 
and we can investigate each Dirac-mode contribution 
to the Polyakov loop individually, based on Eq.(\ref{eq:PLvsD}). 
(For example, each contribution specified by $n$ 
is numerically calculable in lattice QCD.)
In particular, from Eq.(\ref{eq:PLvsD}), 
we can discuss the relation between confinement 
and chiral symmetry breaking in QCD.

As a remarkable fact, because of the factor $\lambda_n^{N_t -1}$, 
the contribution from 
low-lying Dirac-modes with $|\lambda_n|\simeq 0$ 
is negligibly small in the Dirac spectral sum of RHS in Eq.(\ref{eq:PLvsD}),
compared to the other Dirac-mode contribution. 
In fact, the low-lying Dirac modes have tiny contribution 
to the Polyakov loop, regardless of confined or deconfined phase.

%
%

\vspace{0.2cm}
Here, we give several comments on the relation (\ref{eq:PLvsD}) in order.
\vspace{0.2cm} 
\\
1) Equation (\ref{eq:PLvsD}) is a manifestly gauge-invariant relation. 
Actually, the matrix element $\langle n |\hat U_4|n\rangle$ 
can be expressed with 
the Dirac eigenfunction $\psi_n(s)$ and 
the temporal link-variable $U_4(s)$ as 
\begin{eqnarray}
\langle n |\hat U_4|n\rangle =
\sum_s \langle n |s \rangle \langle s 
|\hat U_4| s+\hat t \rangle \langle s+\hat t|n\rangle
=\sum_s \psi_n^\dagger (s)U_4(s) \psi_n(s+\hat t),
\end{eqnarray}
and each term $\psi_n^\dagger (s)U_4(s) \psi_n(s+\hat t)$ 
is manifestly gauge invariant, because of 
the gauge transformation property (\ref{eq:GTDwf}).
[Global phase factors also cancel exactly 
as $e^{-i\varphi_n}e^{i\varphi_n}=1$ 
between $\langle n|$ and $|n \rangle$.] 
\vspace{-0.3cm}
\\
2) In RHS of Eq.(\ref{eq:PLvsD}), 
there is no cancellation between chiral-pair Dirac eigen-states, 
$|n \rangle$ and $\gamma_5|n \rangle$, because $(N_t-1)$ is even, i.e., 
$(-\lambda_n)^{N_t-1}=\lambda_n^{N_t-1}$, and  
$\langle n |\gamma_5 \hat U_4 \gamma_5|n\rangle
=\langle n |\hat U_4|n\rangle$. 
\vspace{0.2cm}
\\
3) Even in the presence of a possible 
multiplicative renormalization factor for the Polyakov loop,
the contribution from the low-lying Dirac modes (or 
the small $|\lambda_n|$ region) is relatively negligible, 
compared to other Dirac-mode contribution 
in the sum of RHS in Eq.(\ref{eq:PLvsD}). 
\vspace{0.2cm}
\\
4) If RHS in Eq.(\ref{eq:PLvsD}) {\it were} not a sum but a product, 
low-lying Dirac modes (or the small $|\lambda_n|$ region) 
should have given an important contribution 
to the Polyakov loop as a crucial reduction factor of $\lambda_n^{N_t-1}$. 
In the sum, however, the contribution ($\propto \lambda_n^{N_t-1}$) 
from the small $|\lambda_n|$ region 
is negligible. 
\vspace{0.2cm}
\\
5) Even if $\langle n |\hat U_4|n\rangle$ behaves as the $\delta$-function 
$\delta(\lambda)$, the factor $\lambda_n^{N_t-1}$ is still crucial 
in RHS of Eq.(\ref{eq:PLvsD}), 
because of $\lambda \delta(\lambda)=0$. 
\vspace{0.2cm}
\\
6) The relation (\ref{eq:PLvsD}) is correct regardless of 
presence or absence of dynamical quarks, 
although the dynamical quark effect appears in $\langle L_P\rangle$, 
the Dirac eigenvalue distribution $\rho(\lambda)$ and 
$\langle n |\hat U_4|n\rangle$.
\vspace{0.2cm}

Note that 
all the above arguments can be numerically confirmed by 
lattice QCD calculations. 
Using actual lattice QCD calculations at the quenched level, 
we numerically confirm the analytical relation (\ref{eq:PLvsD}), 
non-zero finiteness of $\langle n|\hat U_4|n\rangle$, 
and the negligibly small contribution of 
low-lying Dirac modes to the Polyakov loop, 
in both confined and deconfined phases. 

From the analytical relation (\ref{eq:PLvsD}) and the numerical confirmation, 
we conclude that low-lying Dirac-modes 
have tiny contribution to the Polyakov loop, 
and are not essential for confinement, 
while these modes are essential for chiral symmetry breaking.
This conclusion indicates no direct one-to-one correspondence between 
confinement and chiral symmetry breaking in QCD.



\vspace{-0.2cm}
\section*{Acknowledgements}
\vspace{-0.2cm}
H.S. and T.I. are supported in part by the Grant for Scientific Research 
[(C) No.23540306, E01:21105006, No.21674002] 
from the Ministry of Education, Science and Technology of Japan. 

\end{document}